\documentclass[prl,letterpaper,aps,twocolumn]{revtex4}
\usepackage{graphicx}
\usepackage{amsmath}
\begin{document}
\title{Decoherence Dynamics in Low-Dimensional 
Cold Atom Interferometers}
\author{A.A. Burkov, M.D. Lukin and Eugene Demler}
\affiliation{Department of Physics, Harvard University, Cambridge, 
Massachusetts 02138, USA}
\date{\today}
\begin{abstract}
We report on a study of the dynamics of decoherence of a 
matter-wave interferometer, consisting of a pair of low-dimensional 
cold atom condensates at finite temperature. 
We identify two distinct regimes in the time dependence of the coherence 
factor of the interferometer: quantum and classical. 
Explicit analytical results are obtained in both regimes.   
In particular, in the two-dimensional (2D) case in the classical 
(long time) regime, we find that the 
dynamics of decoherence is universal, exhibiting a power-law decay with an 
exponent, proportional to the ratio of the temperature to the 
Kosterlitz-Thouless temperature of a single 2D condensate.
In the one-dimensional (1D) case in the classical regime we find a 
universal nonanalytic time dependence of decoherence,
which is a consequence of the nonhydrodynamic nature of damping 
in 1D liquids.    
\end{abstract}
\maketitle
Understanding quantum phases and phase transitions is a 
substantially more complex task than understanding their classical 
counterparts. 
In particular, in the quantum case equilibrium and nonequilibrium properties 
are inseparable and thus dynamical relaxation 
processes and dissipation may have a profound effect on the nature of 
equilibrium phases and phase transitions.
Superfluid to insulator (SI) transition in disordered 2D superconducting  
films or Josephson junction arrays (JJA)~\cite{Fisher90} 
is probably the most prominent example. 
While it is often thought of as one of the most basic and fundamental 
quantum phase transitions, many aspects of it are still very poorly 
understood. 
In this context it appears interesting to investigate nonequilibrium 
phenomena in 
low-dimensional superfluid systems in a simpler setting 
than is offered by 
such highly complex systems as 
amorphous superconducting films or JJA.
Cold atom systems seem to be especially well-suited for such studies since 
they allow
for an unprecedented control of the relevant experimental parameters 
and for direct, real-time measurements of phase coherence~\cite{Ketterle97,Ketterle04,Polkovnikov06}.   

In this Letter we present a study of the dynamics of decoherence in a
system of two low-dimensional (1D or 2D) cold atom condensates that are 
prepared from a single phase coherent 
condensate by splitting it, using an optical lattice or 
a radio-frequency-field-induced potential on an atom chip. 
It has been recently demonstrated by a number of groups~\cite{Ketterle04} 
that it is possible to split a single condensate in such 
a way that the phase coherence between the two halves is initially well 
preserved. This phase coherent state is, however, a highly 
nonequilibrium one for the split condensate system. 
The system will then relax to thermal equilibrium over time, 
in which the condensates are completely incoherent and it is the dynamics 
of this decoherence process that we are interested in.

The problem of decoherence in cold atom condensates has been studied 
theoretically before by several authors~\cite{Zoller01}. 
Previous studies mostly focused on the {\em single-mode approximation}, 
which assumes that only the lowest-energy phase mode needs to be 
taken into account. 
This approximation is reasonable in three-dimensional (3D) condensates, but 
breaks down in low-dimensional systems, 
where one needs to take into account 
the whole {\em continuum of hydrodynamic (low-energy) modes},
responsible for the enhanced fluctuations~\cite{Shlyapnikov01} and the 
absence of conventional order. 
Quantum decoherence due to such continuum of modes has been explored in a
recent work by Bistritzer and Altman~\cite{Altman06}. 
In this Letter we extend their analysis to investigate both quantum and 
thermal decoherence. 
We also discuss the role of the ``squeezing factor'', determined 
by the finite splitting time. 

At a given temperature $T$ we can divide the hydrodynamic modes 
into two groups: {\em classical}, with energies $\epsilon < T$, and 
{\em quantum}, with $\epsilon > T$. The dynamics of each group of modes 
is governed by classical and quantum fluctuations correspondingly. 
As a consequence, we can also identify two distinct regimes in the 
time dependence of the decoherence process: quantum and classical.   
At short times, i.e. times, shorter than the inverse temperature
$t < 1/T$, decoherence dynamics is dominated by the quantum modes and 
thermal fluctuations may be neglected.  
Since at short times the memory of the initial state of the split system is 
still preserved, the quantum decoherence process is strongly 
influenced by the nature of this initial state, which in turn is determined by 
the process, by which a single condensate is split into two.  
We adopt a simple model, in which during the splitting the system is 
assumed to be described by a Josephson Hamiltonian with a 
time-dependent tunneling term~\cite{Zapata98,endnote1}:
\begin{equation}
\label{eq:8}
H = \frac{E_c}{2} N_r^2  - E_J \cos(\theta) \approx 
\frac{E_c}{2} N_r^2  + \frac{E_J}{2} \theta^2, 
\end{equation}
where     
$N_r = N_1 - N_2$ is the relative number of atoms in the two condensates, 
which is conjugate to the relative phase $[\theta, N_r] = i$. 
The charging energy $E_c = d \mu/ d N = \mu/ N$, where $\mu$ is the 
chemical potential, taken to be the same in both condensates, and   
$E_J$ is the (time-dependent) tunneling energy. 
Eq.(\ref{eq:8}) is the Hamiltonian of a harmonic oscillator with 
a characteristic frequency $\omega = \sqrt{E_c E_J}$ (we will be using 
$\hbar = k_B = 1$ units henceforth).    
One can distinguish two stages of the splitting process~\cite{Leggett98}. 
While $\omega(t) > 1/ \tau_s$, where $\tau_s$ is the characteristic 
time of the splitting, the splitting is approximately adiabatic and 
the ground state 
of Eq.(\ref{eq:8}) should be a good approximation for the actual state of 
the double-well system. Once $\omega(t) < 1 / \tau_s$,
the splitting process is no longer adiabatic. We can then approximate this 
second stage of the splitting process as instantaneous. This is equivalent to 
saying that we approximate the state of the fully split system, when 
$E_J = 0$, by the ground state of Eq.(\ref{eq:8}) at the moment $t^*$ when
the adiabaticity condition fails, i.e. when $\omega(t^*) = 1 / \tau_s$.
Then we obtain the following result for the relative number fluctuations 
of the initial state of the fully split system:
$\Delta N_r \sim \sqrt{N / \mu \tau_s}$,
where $N = \varrho V = (N_1 + N_2)/2$.
This formula agrees well with the apparent value of $\Delta N_r$, which can 
be inferred from the measurements of  
Ref.~\cite{Ketterle06}. The agreement becomes even better if we take 
into account the parabolic trap potential, which 
gives $\mu \sim N^{2/5}$ instead of $\mu \sim N$ for the box geometry. 
It is useful to represent the above result in terms of a 
``squeezing factor'' $\xi$, defined as $1/\xi = \Delta N_r / \sqrt{N} = 
1 / \sqrt{\mu \tau_s}$. Our analysis is valid for $\xi \gg 1$. 
When $\mu \tau_s \lesssim 1$ one should take $\xi \approx 1$.  
Our formula for $\Delta N_r$ thus interpolates between the limits 
of fast splitting ($\tau_s \lesssim 1/\mu$), 
when $\Delta N_r \sim \sqrt{N}$, and 
slow (adiabatic) splitting, in which case $\Delta N_r \sim 0$. 

Choosing the initial state of the system to be a minimum 
uncertainty wavepacket of width $1/\Delta N_r$, 
we find that the coherence factor, defined 
as $\Psi_d(t) = (1/V) \textrm{Re} \int 
d {\bf x} \left\langle e^{i \theta({\bf x}, t)}\right\rangle$, where 
$\theta({\bf x}, t)$ is the relative phase between the two condensates, 
$V = L^d$ is the $d$-dimensional volume of each condensate ($d = 1,2$)  
and the angular brackets denote both quantum and thermal averages, 
is given by: 
\begin{eqnarray}
\label{eq:1}
\Psi_d(t) \propto
\exp[-\mu^2 t^2 / 2 N \xi^2]
\times \left\{ 
\begin{array}{cc}
\exp[- \mu t/ 2 \pi {\cal K} \xi^2], & d = 1, \\
(t/t_0)^{- \mu /16 T_{KT} \xi^2},\,\, & d = 2,
\end{array}
\right.
\end{eqnarray}
where $t_0 \sim 1/\mu$ is 
a short-time cutoff. The Kosterlitz-Thouless temperature $T_{KT}$ of a 
single 2D condensate and the Luttinger parameter ${\cal K}$ in the 1D case 
are defined explicitly below. 
In the fast-splitting regime ($\tau_s \lesssim 1/\mu$) in 1D our result 
agrees with the result of Bistritzer and Altman~\cite{Altman06}. 

At long times $t \gg 1/T$ classical modes, with energies $\epsilon < T$, 
dominate the dynamics. In this case we find:
\begin{eqnarray}
\label{eq:2}
\Psi_d(t) \propto
\left\{ 
\begin{array}{cc}
\exp[-(t/t_0)^{2/3}], & d = 1, \\
(t/t_0)^{-T/8 T_{KT}},\,\, & d = 2,
\end{array}
\right.
\end{eqnarray}
where $t_0$ are (dimension-dependent) cutoff times (time dependences in 
Eq.(\ref{eq:2}) are valid when $t > t_0$), given explicitly below.  
Two features are noteworthy here. In the 1D case classical decoherence dynamics
has a nonanalytic time dependence. As shown below, this is a consequence of 
the fact that in 1D liquids damping at finite temperatures is always 
{\em nonhydrodynamic}, i.e. the damping rate is a nonanalytic function of 
momentum~\cite{Andreev}.
In both 1D and 2D cases the time dependence of decoherence in the classical
regime turns out to be universal,
independent of the microscopic nature of damping and interactions. 
This can make our result especially useful for thermometry in 2D 
condensates.   

We now provide the most important details of the derivation of the above 
results. 
We consider a system of two 1D or 2D superfluids, which are prepared at time 
$t = 0$ in a phase coherent state by, for example, splitting a 
single superfluid in a double-well optical potential. 
This highly nonequilibrium state will then relax to thermal equilibrium. 
We will assume that this relaxation is entirely due to {\em intrinsic 
processes} in each of the two superfluids.
We start from the following imaginary time action for the double-well 
system: 
\begin{eqnarray}
\label{eq:3}  
S&=&\int _0^{1/T} d \tau \int d {\bf x} \left[\Phi^*_{\sigma} \left(
\partial_{\tau} - \mu - \frac {{\boldsymbol \nabla}^2}{2 m} \right) 
\Phi_{\sigma} + \frac{g}{2} |\Phi_{\sigma}|^4\right. \nonumber \\ 
&-&\left. J \Phi_{\sigma}^* 
\tau^x_{\sigma \sigma'} \Phi_{\sigma'}\right].
\end{eqnarray}
Here $\sigma = 1,2$ labels the two condensates (summation over repeated 
indices is implicit) and $J$ is the residual 
tunneling matrix element between the condensates. 
We will assume henceforth that $J$ is negligible and set it to zero.  
Following Popov~\cite{Popov}, we rewrite the action using density-phase 
variables, which we define as $\Phi_{\sigma} = \sqrt{\varrho_{\sigma}} e^{i 
\theta_{\sigma}}$, and expand about the uniform equal density saddle 
point $\varrho_1 = \varrho_2 = \varrho = \mu/g$ 
to obtain the following hydrodynamic imaginary time action:
\begin{eqnarray}
\label{eq:5}
S&=&S_0 + S_1, \nonumber \\
S_0&=&\int_0^{1/T} d \tau \int d {\bf x} \left[i h_{\sigma} \partial_{\tau}
\theta_{\sigma} + \frac{g}{2} h_{\sigma}^2 \right. \nonumber \\ 
&+&\left.\frac{({\boldsymbol \nabla} h_{\sigma})^2}{8 m \varrho} + 
\frac{\varrho ({\boldsymbol \nabla} \theta_{\sigma})^2}{2 m}\right],
\nonumber \\
S_1&=&\int_0^{1/T} d \tau \int d {\bf x} \frac{h_{\sigma} 
({\boldsymbol \nabla} \theta_{\sigma})^2}{2 m},
\end{eqnarray}
where $h_{\sigma} = \varrho_{\sigma} - \varrho$.
The saddle point expansion is legitimate at sufficiently low temperatures, 
such that the (exponentially small) contribution of phase slips (1D) or 
vortices (2D) to 
correlations can be neglected.   
The harmonic part of the action $S_0$ describes undamped collective 
density-phase modes of the double-well superfluid. 
The anharmonic part of the action $S_1$ is responsible for the dissipation 
and relaxation to equilibrium at finite temperatures, as will be shown 
below~\cite{Martin,Popov}.
As discussed above, at short times $t < 1/T$ after the decoupling of the two 
condensates, the contribution of thermal fluctuations of the relative phase 
to decoherence dynamics is negligible and  
temperature can simply be set to zero:   
the dynamics of decoherence in this case is purely quantum. 
Furthermore, we may neglect the anharmonic terms in Eq.(\ref{eq:5}), which 
in the quantum regime are unimportant. 
Introducing the relative and center-of-mass phase variables
$\theta = \theta_1 - \theta_2, \,\chi = (\theta_1 + \theta_2)/2$ 
and integrating over density fluctuations, we obtain:
\begin{eqnarray}
\label{eq:6}
S&=&S_{\chi} + S_{\theta}, \nonumber \\
S_{\chi}&=& \int_0^{1/T} d \tau \int d {\bf x} \left[\frac{1}{g} (\partial_{\tau} 
\chi)^2 + \frac{\varrho}{m} ({\boldsymbol \nabla} \chi)^2 \right], 
\nonumber \\
S_{\theta}&=& \int_0^{1/T} d \tau \int d {\bf x} \left[\frac{1}{4 g} (\partial_{\tau} 
\theta)^2 + \frac{\varrho}{4 m} ({\boldsymbol \nabla} \theta)^2 \right].
\end{eqnarray}
The relative and center-of-mass phase dynamics are thus completely decoupled
at short times.
As the action for the relative phase is purely harmonic, the problem of the 
relative phase dynamics at short times may be solved exactly~\cite{Altman06}.
Passing from the imaginary time action for the long-wavelength relative 
phase modes Eq.(\ref{eq:6}) back to the Hamiltonian, and rewriting it in 
Fourier space, we obtain:
\begin{equation}
\label{eq:7}
H = \sum_{\bf k} \left[ g |\Pi_{\bf k}|^2 + \frac{\varrho {\bf k}^2}{4 m} 
|\theta_{\bf k}|^2 \right], 
\end{equation}
where $[\theta_{\bf k}, \Pi_{{\bf k}'}] = i \delta_{{\bf k} {\bf k}'}$.  

Taking the initial state of the split-condensate system $|\Phi_0\rangle$ 
to be a minimum uncertainty wavepacket, as discussed above, 
it is straightforward to evaluate the time evolution of the 
coherence factor~\cite{Altman06}. One obtains:
\begin{eqnarray}
\label{eq:9}
\Psi(t)&=&\frac{1}{V} \textrm{Re} \int d {\bf x} \langle \Phi_0 |
e^{i H t}
e^{i \theta({\bf x})} e^{- i H t}|\Phi_0 \rangle \nonumber \\
&=&e^{-(1/ 2 V) \sum_{\bf k} \langle |\theta_{\bf k}|^2 \rangle_t},
\end{eqnarray} 
where
\begin{eqnarray}
\label{eq:10}
\langle |\theta_{\bf k}|^2 \rangle_t &\equiv& 
\langle \Phi_0 | e^{i H t} |\theta_{\bf k}|^2 e^{- i H t} 
|\Phi_0 \rangle \nonumber \\ 
&=&\frac{\mu \tau_s}{\varrho}\left[ 
\cos^2(\epsilon_{\bf k} t) + \frac{1}{\epsilon_{\bf k}^2 \tau_s^2} 
\sin^2(\epsilon_{\bf k} t)\right],
\end{eqnarray} 
and $\epsilon_{\bf k} = \sqrt{g \varrho/m} k \equiv c_s k$.  
Evaluating Eq.(\ref{eq:9}) at times $t > 1/\mu$, we obtain 
Eq.(\ref{eq:1}), with $T_{KT} = \pi \varrho/2 m$ and
${\cal K} = (\pi/2) \sqrt{\varrho/g m}$ (these are weak-coupling expressions
for $T_{KT}$ and ${\cal K}$, which are expected to be accurate at low 
temperatures). 
Note that in Eq.(\ref{eq:1}) we have explicitly separated 
the contribution 
of the ${\bf k} = 0$ mode (the common factor), which will contribute 
separately from the $ k > 0$ continuum in a finite-size system.
This is nothing 
but the ``phase diffusion''~\cite{Ketterle06}, which is the only contribution
that exists at long times in bulk systems.     
 
We now extend the above theory to the classical, i.e. long time 
$t > 1/T$ limit. 
The character of the long time dynamics is determined by the low-energy, 
classical modes, with energies $\epsilon < T$. 
We start from the hydrodynamic action for density and phase fluctuations 
Eq.(\ref{eq:5}).
Unlike in the quantum case, the anharmonic terms in the action turn out to be 
crucial, as will become clear below.
Despite the fact that the source of relaxation is purely intrinsic 
in our case, it turns out to be possible, and very useful, to cast the 
problem into the form~\cite{Leggett81} of an ``observable macroscopic 
variable'' interacting with a ``thermal bath'' of microscopic degress of 
freedom.
This is possible thanks to the following observation. 
While in the imaginary time action Eq.(\ref{eq:5}) all degrees 
of freedom enter on a completely equal footing, 
the initial conditions for the relative and center-of-mass variables, 
introduced above,  are drastically different.
Indeed, we can formally model the process of splitting a single condensate 
into two by suddenly changing the value of the tunneling amplitude in 
Eq.(\ref{eq:1}) from a large value at times $t < 0$ to zero at $t = 0$.
In hydrodynamic description this is equivalent to suddenly driving 
a large gap for the relative phase-density collective modes to zero. 
Center-of-mass modes however, feel this change only weakly, through 
anharmonic terms in the hydrodynamic action. 
It thus seems reasonable to assume that the center-of-mass modes 
approximately 
remain in thermal equilibrium throughout the splitting and subsequent 
relaxation process~\cite{endnote2}. 
We can then think of these degrees of freedom as forming a thermal bath.
The relative density and phase modes, in contrast, are strongly affected 
by the separation 
process and are far out of equilibrium at $t = 0$. 
We thus arrive at the picture of out-of-equilibrium relative phase collective 
modes, 
interacting with a thermal bath of the center-of-mass modes, the interaction 
being 
described by the anharmonic terms in Eq.(\ref{eq:5}).
To make this separation of the degrees of freedom explicit we can 
perturbatively integrate out the center-of-mass degrees of freedom (we expect 
this perturbation theory to work well at long times)
in Eq.(\ref{eq:5}), 
and obtain the following effective harmonic action for the relative phase 
variables only:
\begin{eqnarray}
\label{eq:11}
&&S = \int_0^{1/T} d \tau \sum_{\bf k} \left[ 
\frac{1}{4 g} |\partial_{\tau} \theta({\bf k},\tau)|^2 + 
\frac{\varrho {\bf k}^2}{4 m}
|\theta({\bf k},\tau)|^2 \right] \nonumber \\ 
&+& \int_0^{1/T} d \tau_1 d \tau_2 \sum_{\bf k} 
\Pi({\bf k}, \tau_1 - \tau_2) 
\theta({\bf k}, \tau_1) \theta(-{\bf k}, \tau_2).
\end{eqnarray}
In 2D the lowest order (single phonon bubble) approximation for the 
dissipative kernel $\Pi({\bf k}, i \omega)$ is 
sufficient~\cite{Popov,Belyaev}.  
In 1D the contribution of a single bubble diagram diverges ``on-shell'', 
i.e. at $i \omega = \epsilon_{\bf k}$, which is a consequence of kinematics 
in 1D, namely the fact that the laws of energy and momentum conservation are 
satisfied simultaneously~\cite{Andreev}. It is then necessary to resum the 
most divergent (maximal number of bubbles) diagrams at each 
order~\cite{endnote3}.
We obtain:   
\begin{eqnarray}
\label{eq:12}
\Pi({\bf k}, i\omega) = \frac{|\omega|}{8 g} \times \left\{
\begin{array}{cc}
\gamma_1 \epsilon_{\bf k}^{3/2},\, & d = 1, \\
\gamma_2 \epsilon_{\bf k},\, & d = 2,
\end{array}
\right.
\end{eqnarray}
where $\gamma_1 = \sqrt{2 \alpha T {\cal K}}/ \pi \mu$ 
($\alpha \approx 1.954$ is a numerical constant) and $\gamma_2 = 4 \pi T^2 /
3 \sqrt{3} \mu T_{KT}$ are damping coefficients, 
characterizing the strength of dissipation. Note that while $\gamma_2$ is 
dimensionless, $\gamma_1$ has dimensions of $1/\sqrt{\epsilon}$. 
The most interesting feature of Eq.(\ref{eq:12}) is the nonanalytic 
dependence of $\Pi({\bf k}, i \omega)$ on ${\bf k}$ in the 
1D case~\cite{Andreev}, which was first noted 
in a different context by Andreev. This means that damping 
in a 1D liquid at finite
temperatures is {\em nonhydrodynamic}, which is a consequence of the breakdown 
of superfluid order in 1D on length scales longer than the 
temperature-dependent correlation length~\cite{Andreev}. 
It is this nonanalytic dependence of 
the damping kernel on momentum that leads to the nonanalytic time 
dependence of decoherence in Eq.(\ref{eq:2}). 

Using Keldysh formalism, it is possible to show that 
the imaginary time action Eq.(\ref{eq:11}) is {\em exactly} 
equivalent to the following real time quantum Langevin 
equation~\cite{Schmid82,Sachdev96}:
\begin{eqnarray}
\label{eq:13}
&&\frac{d^2 \theta({\bf k}, t)}{d t^2} + 
\epsilon_{\bf k}^2 \theta({\bf k}, t)
+ 8 g \int_{-\infty}^t d t' 
\Pi_I({\bf k}, t - t') \theta({\bf k}, t') \nonumber \\ 
&=&2 g \zeta({\bf k}, t), 
\end{eqnarray}
where the quantum noise variable $\zeta({\bf k}, t)$ is defined by its 
autocorrelation function $\langle \zeta({\bf k}, t_1) \zeta(-{\bf k}, t_2) 
\rangle = 2 \Pi_R({\bf k}, t_1 - t_2)$ and the functions $\Pi_{R,I}$ are 
given by:
\begin{eqnarray}
\label{eq:14}
\Pi_I({\bf k}, \omega)&=&i\,\, \textrm{Im} \Pi({\bf k}, i\omega \rightarrow
\omega + i 0+), \nonumber \\
\Pi_R({\bf k}, \omega)&=&i\,\, \Pi_I({\bf k}, \omega) \coth(\omega/ 2 T). 
\end{eqnarray}
Solving Eq.(\ref{eq:13}) with the initial conditions $\theta({\bf k},0) = 0,\,
\dot \theta({\bf k}, 0) = 0$~\cite{endnote4}, by the Laplace transform, 
we obtain:
\begin{equation}
\label{eq:15}
\langle \theta^2(\epsilon, t) \rangle_d = \frac{T \mu}{\varrho \epsilon^2}
[1 - f_d(\epsilon,t)], 
\end{equation}
where
\begin{eqnarray}
\label{eq:16}
f_1(\epsilon,t)&=&
e^{-\gamma_1 \epsilon^{3/2} t} \left[1 - \frac{\gamma_1^2 \epsilon}{2} 
\sin^2(\epsilon t)
+ \frac{\gamma_1 \sqrt{\epsilon}}{2} \sin(2 \epsilon t)\right], \nonumber \\
f_2(\epsilon,t)&=&
e^{-\gamma_2 \epsilon t}\left[1 - \frac{\gamma_2^2}{2} \sin^2(\epsilon t)
+ \frac{\gamma_2}{2} \sin(2 \epsilon t)\right].
\end{eqnarray}
Note that in the limit $t \rightarrow \infty$ Eq.(\ref{eq:15}) 
correctly reproduces the 
equilibrium magnitude of thermodynamic fluctuations of $\theta(\epsilon)$.
Now we can easily evaluate the time dependence of the coherence factor. 
At long times we obtain:
\begin{equation}
\label{eq:17}
\Psi_d(t) = \exp \left[- \frac{1}{2} \int_0^T d \epsilon 
\,\,\nu_d(\epsilon) \langle \theta^2(\epsilon,t) \rangle_d \right],
\end{equation}
where $\nu_d(\epsilon)$ is the phonon density of states in a $d$-dimensional
superfluid.  
Evaluating the integral over $\epsilon$, we arrive at 
Eq.(\ref{eq:2}), with the cutoff times given by:
\begin{eqnarray}
\label{eq:18}
t_0 = \left\{
\begin{array}{cc} 
\beta \pi \mu {\cal K}/ T^2,\, & d = 1,\\
1/\gamma_2 T,\,& d = 2,
\end{array}
\right.
\end{eqnarray}
where $\beta = [8 / (2 \alpha)^{1/3} \Gamma(1/3)]^{3/2} \approx 2.61$.
   
\acknowledgments{We acknowledge useful discussions with I. Affleck, E. Altman,
I. Bloch, V. Gritsev, B.I. Halperin, W. Ketterle, M. Oberthaler, 
A. Polkovnikov, S. Sachdev, and J. Schmiedmayer.  
Financial support was provided by the National Science Foundation 
under grants DMR05-41988 and DMR01-32874, by the Harvard-MIT CUA, and AFOSR.}


\begin{thebibliography}{99}
\bibitem{endnote1} This analysis is done for the ${\bf k} = 0$ mode. 
One can apply the same arguments for all modes with no change 
in the final result.
\bibitem{endnote2} In reality some degree of heating of these modes 
will occur. One should then think of the temperature $T$ in 
our equations as the temperature of the final equilibrium 
state of the system.
\bibitem{endnote4} We expect them to be a good 
approximation to the exact initial conditions in the long time limit.
\end{thebibliography}
\end{document}